\documentclass[amsmath,amssymb,12pt,superscriptaddress,nofootinbib]{revtex4}

\usepackage{graphicx}
\usepackage{dcolumn}
\usepackage{bm}

\newcommand{\dd}{\mathrm{d}}

\newcommand{\del}{\partial}
\newcommand{\ee}{{\rm e}}

\begin{document}
\baselineskip5.5mm
\thispagestyle{empty}

{\baselineskip0pt
\leftline{\baselineskip14pt\sl\vbox to0pt{
               \hbox{\it Yukawa Institute for Theoretical Physics} 
              \hbox{\it Kyoto University}
               \vss}}
\rightline{\baselineskip16pt\rm\vbox to20pt{
            \hbox{YITP-12-63}
\vss}}%
}

\author{Chul-Moon Yoo}\email{yoo@yukawa.kyoto-u.ac.jp}
\affiliation{
Yukawa Institute for Theoretical Physics, Kyoto University, 
Kyoto 606-8502, Japan
}

\vskip2cm
\title{Notes on Spinoptics in a Stationary Spacetime
}
  
\begin{abstract}
In Ref.~\cite{Frolov:2011mh}, 
equations of the modified 
geometrical optics for circularly polarized photon trajectories 
in a stationary spacetime 
are derived by using a (1+3)-decomposed form of Maxwell's equations. 
We derive the same results by using a 
four-dimensional covariant description. 
In our procedure, the null nature of the modified photon trajectory 
naturally appears and the energy flux is apparently null. 
We find that, in contrast to the standard 
geometrical optics, 
the inner product of the stationary Killing vector 
and the tangent null vector to 
the modified photon trajectory is no longer a
conserved quantity along light paths. 
This quantity is furthermore different for left and right handed photon. 
A similar analysis is performed for gravitational waves and 
an additional factor of 2 appears 
in the modification due to the spin-2 nature of gravitational waves. 
\end{abstract}

\maketitle
\pagebreak

%%%%%%%%%%%%%%%%%%%%%%%%%%%%%%%%%%%%%%%%%%%%%%%%%%%%%%%%%%%%%%%%
\section{introduction}
%%%%%%%%%%%%%%%%%%%%%%%%%%%%%%%%%%%%%%%%%%%%%%%%%%%%%%%%%%%%%%%%
Light propagation in the gravitational field 
of a rotating body has been a topic of study in the past several years. 
One phenomenon of interest is 
rotation of the polarization vector, known as 
gravitational 
Faraday rotation~\cite{Balazs:1958zz,Plebanski:1959ff,1982GReGr..14..865F,
Ishihara:1987dv,1992PhRvD..46.5407C,NouriZonoz:1999pb,Sereno:2004jx}. 
This effect does not manifest in the gravitational field 
of a non-rotating body, 
such as Schwarzschild spacetime~\cite{Mashhoon:1973zz}, 
while it does occur
%MOD%
 for Kerr spacetime. 
This fact suggests that the existence of 
helicity--rotation coupling and 
the propagation of circularly polarized 
electro-magnetic waves depends on the helicity. 
Its occurrence has been confirmed 
by analyzing Maxwell's equations in curved spacetimes 
created by rotating bodies~\cite{Mashhoon:1974cq,Mashhoon:1974,Mashhoon:1975ki}. 
This effect is also discussed based on the 
gravitational Larmor's theorem~\cite{1993PhLA..173..347M,Ramos:2006sb}. 

In Ref.~\cite{Frolov:2011mh}, Frolov and Shoom reported that 
the spinoptics in a gravitational field created by a 
rotating massive compact object can be described by 
a modified geometrical optics approximation. 
They used a (1+3)-decomposed form of Maxwell's equations 
and also considered a standard geometrical optics approach. 
In their setting of the base vector field 
for the circular polarization, 
a phase shift appears that depends on the helicity. 
They 
proposed a modification 
in which the ordering of the equations associated with the geometrical 
optics approximation 
is changed so that the phase shift is absorbed in the eikonal of the eikonal ansatz. 
This treatment also 
leads to a modification of the photon trajectory 
depending on the helicity. 
Using this procedure, scattering of circularly polarized 
light by a rotating black hole is discussed in Ref.~\cite{Frolov:2012zn}. 

In this paper, we derive the same 
equations of the modified geometrical optics as in Ref.\cite{Frolov:2011mh} 
by using another description in which 
four-dimensional covariance is maintained. 
In our procedure, we can easily see the four-dimensional picture 
of the photon propagation. 
In addition, this procedure can be easily applied to the case of 
gravitational waves, as will be explicitly shown(see also Ref.~\cite{Ramos:2006sb}). 

This paper is organized as follows. 
In Sec.~\ref{sec:TTE}, 
we review the standard geometrical optics approximation 
for Maxwell's equations in a stationary spacetime. 
We introduce a circular polarization base vector field 
in Sec.~\ref{sec:basevec} 
using an identical method to that in Ref.~\cite{Frolov:2011mh}. 
We then discuss the transport equation for the polarization vector 
using the circular polarization base vector 
field in Sec.~\ref{sec:PTPV}. 
The photon trajectory, transport equation and 
energy flux in the modified geometrical optics 
are given in Sec.~\ref{sec:MGO}. 
In Sec.~\ref{sec:GS} the procedure is applied to the case of gravitational waves. 
Sec.~\ref{sec:SD} is devoted to a summary and discussion. 

%%%%%%%%%%%%%%%%%%%%%%%%%%%%%%%%%%%%%%%%%%%%%%%%%%%%%%%%%%%%%%%%
\section{Standard Geometrical Optics}
\label{sec:sgo}
%%%%%%%%%%%%%%%%%%%%%%%%%%%%%%%%%%%%%%%%%%%%%%%%%%%%%%%%%%%%%%%%
%%%%%%%%%%%%%%%%%%%%%%%%%%%%%%%%%%%%%%%%%%%%%%%%%%%%%%%%%%%%%%%%
\subsection{Trajectory, Transport Equation and Energy Flux}
\label{sec:TTE}
%%%%%%%%%%%%%%%%%%%%%%%%%%%%%%%%%%%%%%%%%%%%%%%%%%%%%%%%%%%%%%%%
In this paper 
we focus on a stationary spacetime manifold $(\mathcal M, g)$, 
where $\mathcal M$ is a four-dimensional manifold 
with a smooth Lorentzian metric $g$ which has 
a smooth 1-parameter group $G$ of the 
isometry generated by the Killing vector field $\xi$. 
Following Ref.~\cite{Frolov:2011mh}, 
we write the line element of the stationary spacetime as 
follows:
\begin{equation}
\dd s^2= -h(\dd t-\hat g_i\dd x^i)^2+h\hat \gamma_{ij}\dd x^i \dd x^j,  
\end{equation}
where $i$, $j$ run from 1 to 3 and $h$, $\hat g_i$ and $\hat \gamma_{ij}$
are functions of $x^i$. 
The stationary Killing vector field is given by  
\begin{equation}
\xi=\frac{\del}{\del t},~~\xi^\mu\xi_\mu=-h<0. 
\end{equation}
Using the action of the isometry group $G$ on $\mathcal M$, 
we can define the orbit space associated with 
the Killing vector field $\xi$ as $\mathcal N:=\mathcal M/G$. 
We define the normalized Killing vector field $u$ by 
\begin{equation}
u^\mu:=\xi^\mu / \sqrt{h}. 
\end{equation}
For later convenience, we define the projection tensor 
$\gamma$ by 
\begin{equation}
\gamma_{\mu\nu}:=g_{\mu\nu}+u_\mu u_\nu. 
\end{equation}
Then, $\gamma_{ij}=h\hat \gamma_{ij}$ and $\gamma$ 
gives the naturally induced metric on $\mathcal N$. 

Note that in this paper, we consider the region in which $h$ is positive definite. 
This condition may not be satisfied for regions within the ergosphere 
of a Kerr black hole. 
Therefore, as with the formalism in Refs.~\cite{Frolov:2011mh,Frolov:2012zn}, 
our formalism cannot be straightforwardly applied to the ergoregion 
with the Killing vector field which is tangent to the world line of 
the static observer at the infinity.

In Ref.~\cite{Frolov:2011mh}, 
Maxwell's equations 
are reduced to the master equations on 
the orbit space $\mathcal N$ with the metric $\hat \gamma$. 
We do not follow the same procedure and 
instead use the four-dimensional covariant form of the equations. 
We consider the vector potential $A_\mu$ 
which satisfies the Lorenz gauge condition given by 
\begin{equation}
\nabla_\mu A^\mu=0
\label{eq:lorenz1}
\end{equation}
and 
the wave equation given by
\begin{equation}
\nabla^\nu\nabla_\nu A_\mu-R_\mu^{~\nu} A_\nu=0, 
\label{eq:waveeq1}
\end{equation}
where $R_{\mu\nu}$ is the Ricci curvature tensor.
Following the standard method (e.g. Refs.~\cite{1967ZNatA..22.1328E,Misner:1974qy}), 
we write the eikonal ansatz as follows:
\begin{equation}
A_\mu=\left(a_\mu+\epsilon b_\mu+\mathcal O(\epsilon^2)\right)e^{iS/\epsilon}, 
\label{eq:ansatz}
\end{equation}
where $\epsilon$ is a book-keeping parameter 
%MOD%
that we take to be small during our
manipulations; at the end of our calculations 
we reset it to $\epsilon\rightarrow1$, so that S
becomes the actual phase function.
%MOD%

Substituting the ansatz~\eqref{eq:ansatz} into Eq.~\eqref{eq:lorenz1}, 
we obtain the following equation from the order of $\epsilon^{-1}$:
\begin{equation}
a^\mu k_\mu=0, 
\label{eq:normality1}
\end{equation}
where $k_\mu$ is defined by 
\begin{equation}
k_\mu:=\nabla_\mu S. 
\label{eq:kdef}
\end{equation}
From the order of $\epsilon^{-2}$ in Eq.~\eqref{eq:waveeq1}, 
we obtain 
\begin{equation}
k^\mu k_\mu=0. 
\label{eq:null1}
\end{equation}
Rewriting Eq.~\eqref{eq:null1} as 
\begin{equation}
\mathcal H:=\frac{1}{2}g^{\mu\nu}\nabla_\mu S \nabla_\nu S=0, 
\label{eq:hamija1}
\end{equation}
we can regard this equation as a 
Hamilton--Jacobi equation for $S$. 
Since the four-velocity of the corresponding dynamical system to 
Eq.~\eqref{eq:hamija1} is given by $k^\mu$, 
we can regard the Hamiltonian equation for 
the Hamiltonian \eqref{eq:hamija1} 
as the equation for the ray trajectory generated by $k^\mu$. 
The Hamiltonian equations are given by 
\begin{equation}
k^\nu\nabla_\nu k^\mu=0. 
\label{eq:nullgeo}
\end{equation}
Hence trajectories are given by null geodesics. 
This equation can be simply derived by 
differentiating Eq.~\eqref{eq:null1} and using Eq.~\eqref{eq:kdef}.

The order of $\epsilon^{-1}$ in Eq.~\eqref{eq:waveeq1} 
gives the following transport equation:
\begin{equation}
k^\nu\nabla_\nu a^\mu+\frac{1}{2}a^\mu\nabla_\nu k^\nu=0. 
\label{eq:transport1}
\end{equation}
Following convention, 
we divide $a^\mu$ into the real amplitude $a$ and 
the complex polarization vector $\ell^\mu$ as follows:
\begin{equation}
a_\mu=a\ell_\mu~,~~\ell^\mu \overline \ell_\mu=1~,~~a\in \mathbb R, 
\label{eq:divide1}
\end{equation}
where $\overline \ell_\mu$ denotes the complex conjugate of $\ell_\mu$. 
Then, contracting $\overline \ell_\mu$ with the transport equation 
\eqref{eq:transport1}, 
from the real part, 
we obtain
\begin{equation}
\nabla_\mu(a^2k^\mu)=0. 
\label{eq:phconv1}
\end{equation}
Substituting this equation into \eqref{eq:transport1}, 
we obtain  
\begin{equation}
k^\nu\nabla_\nu \ell^\mu=0. 
\label{eq:paratr}
\end{equation}
Eq.~\eqref{eq:phconv1} describes the conservation of the photon number 
and Eq.~\eqref{eq:paratr} indicates that 
the polarization vector $\ell^\mu$ is parallel-transported 
along the ray trajectory.

The field strength $F_{\mu\nu}$ of 
the vector potential \eqref{eq:ansatz} 
is given by 
\begin{equation}
F_{\mu\nu}={\rm Re}\left\{\nabla_\mu A_\nu-\nabla_\nu A_\mu\right\}
\simeq2a {\rm Re}\left\{i\ee^{iS}k_{[\mu}\ell_{\nu]}\right\}
\label{eq:field}
\end{equation}
at the leading order of the geometrical optics approximation, 
where square brackets denote anti-symmetrization. 
Then, the energy momentum tensor $T^{\mu\nu}$ is given by
\begin{eqnarray}
T^{\mu\nu}&=&\frac{1}{4\pi}\left(F^\mu_{~\lambda}F^{\nu\lambda}
-\frac{1}{4}g^{\mu\nu}F_{\lambda\sigma}F^{\lambda\sigma}\right)\cr
&=&\frac{a^2}{8\pi}k^\mu k^\nu
\left(1-{\rm Re}\left\{\ee^{2iS}\ell_\lambda\ell^\lambda\right\}\right). 
\end{eqnarray}
Averaging over several wavelengths, we obtain
\begin{equation}
\left<T^{\mu\nu}\right>=\frac{a^2}{8\pi}k^\mu k^\nu. 
\end{equation}
This expression indicates that the 
energy flux is proportional to $k^\mu$ and null
at the leading order of the standard geometrical optics approximation.

%%%%%%%%%%%%%%%%%%%%%%%%%%%%%%%%%%%%%%%%%%%%%%%%%%%%%%%%%%%%%%%%
\subsection{Base Vector Fields}
\label{sec:basevec}
%%%%%%%%%%%%%%%%%%%%%%%%%%%%%%%%%%%%%%%%%%%%%%%%%%%%%%%%%%%%%%%%
Taking stationarity into account, we impose 
\begin{equation}
\mathcal L_\xi k^\mu=\xi^\nu\nabla_\nu k^\mu-k^\nu\nabla_\nu \xi^\mu=0, 
\label{eq:liek}
\end{equation}
where $\mathcal L_\xi$ is the Lie derivative with respect to $\xi$. 
Using this equation and $\nabla_\mu k_\nu=\nabla_\nu k_\mu$, 
we obtain 
\begin{equation}
\nabla_\mu(\xi^\nu k_\nu)=0. 
\end{equation}
We define the frequency $\omega$ as follows:
\begin{equation}
\omega:=-\xi^\mu k_\mu. 
\label{eq:defomega}
\end{equation}

We introduce the spacelike unit vector 
along the ray direction $n^\mu$, given by 
\begin{equation}
n^\mu:=\frac{\sqrt{h}}{\omega}k^\mu-u^\mu. 
\label{eq:n2k}
\end{equation}
This satisfies
\begin{equation}
n^\mu n_\mu=1 ,~~n^\mu u_\mu=0. 
\end{equation}
To set an orthonormal base system, we 
define two additional unit spacelike vector fields $e^\mu_1$ and $e^\mu_2$, given below. 
First, at a point, we set $e^\mu_A$ such that 
the following conditions are satisfied:
\begin{equation}
g_{\mu\nu}e^\mu_Ae^\nu_B=\delta_{AB}~,~~u_\mu e^\mu_A=n_\mu e^\mu_A=0, 
\label{eq:orthonormal}
\end{equation}
where $A=1$, 2. 
Then, following Ref.~\cite{Frolov:2011mh}, 
we extend $e^\mu_A$ along the integral curve of $n^\mu$ 
by imposing the following condition: 
\begin{eqnarray}
\mathcal F_n e^\mu_A&:=&n^\nu D_\nu e^\mu_A+e^\nu_A(n^\lambda D_\lambda n_\nu)n^\mu
-(e^\nu_An_\nu)n^\lambda D_\lambda n^\mu=0\cr
&&\Leftrightarrow n^\nu D_\nu e^\mu_A=-e^{\nu}_{A}(n^\lambda D_{\lambda} n_{\nu})n^\mu
=n^\mu n_\nu n^\lambda D_\lambda e^\nu_A, 
\label{eq:Fermi}
\end{eqnarray}
where the action of 
$D_\mu$ on a vector field $v^\mu$ is defined by 
\begin{equation}
D_\mu v_\nu=\gamma_\mu^{~\rho}\gamma_\nu^{~\lambda}\nabla_\rho v_\lambda. 
\end{equation}
We can check that the condition Eq.~\eqref{eq:Fermi} is 
equivalent to Eq.~(85) in Ref.~\cite{Frolov:2011mh}, 
which gives Fermi transport on $(\mathcal N,\hat\gamma)$. 
In addition, we extend the base vector fields 
along the integral curve of $\xi$ by 
the Lie transport as follows:
\begin{equation}
\mathcal L_\xi e^\mu_A=0. 
\end{equation}
Then, Eq.~\eqref{eq:orthonormal} is satisfied at any point of 
the spacetime. 

Finally, we define the circular polarization base vector field as follows:
\begin{equation}
m^\mu=(e^\mu_1+i\sigma e^\mu_2)/\sqrt{2}, 
\end{equation}
where $\sigma=\pm1$ specifies circular 
polarization. 
Then, we have 
\begin{eqnarray}
e^\mu_1&=&\sqrt{2}(m^\mu+\overline m^\mu), \\
e^\mu_2&=&-i\sigma\sqrt{2}(m^\mu-\overline m^\mu). 
\end{eqnarray}
$m^\mu$ and $\overline m^\mu$  also 
satisfy Eq.~\eqref{eq:Fermi} and are Lie transported 
along the $\xi$ direction.

%%%%%%%%%%%%%%%%%%%%%%%%%%%%%%%%%%%%%%%%%%%%%%%%%%%%%%%%%%%%%%%%
\subsection{Parallel Transport of the Polarization Vector}
\label{sec:PTPV}
%%%%%%%%%%%%%%%%%%%%%%%%%%%%%%%%%%%%%%%%%%%%%%%%%%%%%%%%%%%%%%%%

As shown in Eq.~\eqref{eq:paratr}, 
the polarization vector $\ell^\mu$ is parallel-transported 
along the null geodesic generated by $k^\mu$. 
Using the circular polarization base vector field $m^\mu$, we can write 
\begin{equation}
\ell^\mu=m^\mu \ee^{i\varphi}, 
\label{eq:defvarphi}
\end{equation}
where $\varphi$ is a real function of $x^i$. 
Then, Eq.~\eqref{eq:paratr} can be rewritten as 
\begin{equation}
k^\nu\nabla_\nu(m^\mu\ee^{i\varphi})=0
\Leftrightarrow
m^\mu k^\nu\nabla_\nu (\ee^{i\varphi})=-\ee^{i\varphi}k^\nu\nabla_\nu m^\mu. 
\end{equation}
Contracting with $\overline m^\mu$, 
we obtain
\begin{eqnarray}
ik^\nu \nabla_\nu \varphi &=&
m^\mu k^\nu\nabla_\nu \overline m_\mu. 
\end{eqnarray}
Using \eqref{eq:n2k}, we find 
\begin{eqnarray}
ik^\nu \nabla_\nu \varphi &=&
\frac{\omega}{\sqrt{h}} m^\mu (n^\nu+u^\nu)\nabla_\nu \overline m_\mu\cr
&=&\frac{\omega}{\sqrt{h}} m^\mu n^\nu \nabla_\nu \overline m_\mu
+\frac{\omega}{h}m^\mu \xi^\nu \nabla_\nu \overline m_\mu\cr
&=&\frac{\omega}{\sqrt{h}} m^\mu n^\nu D_\nu \overline m_\mu
+\frac{\omega}{h}m^\mu \xi^\nu \nabla_\nu \overline m_\mu\cr
&=&\frac{\omega}{h}m^\mu \xi^\nu \nabla_\nu \overline m_\mu\cr
&=&\frac{\omega}{h}m^\mu \overline m^\nu \nabla_\nu \xi_\mu, 
\label{eq:kdphi1}
\end{eqnarray}
where we have used $\mathcal F_n \overline m^\mu=0$ and 
$\mathcal L_\xi \overline m^\mu=0$. 
Since $\nabla_\mu\xi_\nu$ is anti-symmetric, 
we obtain 
\begin{eqnarray}
k^\nu \nabla_\nu \varphi 
&=&
\sigma\frac{\omega}{h} \ee^{[\mu}_1\ee^{\nu]}_2\nabla_\nu \xi_\mu\cr
&=&\frac{1}{2}
\sigma\frac{\omega}{h} 
u_\rho n_\lambda \varepsilon^{\mu\nu\rho\lambda} \nabla_\nu \xi_\mu\cr
&=&\frac{1}{2}
\sigma u_\rho k_\lambda \varepsilon^{\mu\nu\rho\lambda} \nabla_\nu u_\mu, 
\label{eq:kdphi2}
\end{eqnarray}
where $\varepsilon^{\mu\nu\rho\lambda}$ is the completely anti-symmetric tensor 
with $\varepsilon^{0123}=1/\sqrt{-\det g}$. 
Performing (1+3) decomposition, we can check that 
Eq.~\eqref{eq:kdphi2} is equivalent to Eq.~(102) in Ref.~\cite{Frolov:2011mh}. 
This is the well-known gravitational analogue of the Faraday rotation~\cite{Balazs:1958zz,Plebanski:1959ff,1982GReGr..14..865F,
Ishihara:1987dv,1992PhRvD..46.5407C,NouriZonoz:1999pb,Sereno:2004jx}.

%%%%%%%%%%%%%%%%%%%%%%%%%%%%%%%%%%%%%%%%%%%%%%%%%%%%%%%%%%%%%%%%
\section{Modified Geometrical Optics}
\label{sec:MGO}
%%%%%%%%%%%%%%%%%%%%%%%%%%%%%%%%%%%%%%%%%%%%%%%%%%%%%%%%%%%%%%%%

%%%%%%%%%%%%%%%%%%%%%%%%%%%%%%%%%%%%%%%%%%%%%%%%%%%%%%%%%%%%%%%%
\subsection{Modification of the Eikonal}
\label{sec:mei}
%%%%%%%%%%%%%%%%%%%%%%%%%%%%%%%%%%%%%%%%%%%%%%%%%%%%%%%%%%%%%%%%

The guiding principle of the 
modification is that 
the phase term $\ee^{i\varphi}$ should be included in 
the eikonal(see Eqs.~\eqref{eq:ansatz}, \eqref{eq:divide1}, and \eqref{eq:defvarphi}), that is, 
the modified eikonal $\widetilde S$ should be given by 
\begin{equation}
S\rightarrow \widetilde S \sim S+\varphi. 
\label{eq:tildes}
\end{equation}
Before modification, the Hamiltonian of the ray trajectory 
is given by 
\begin{equation}
\mathcal H=\frac{1}{2}g^{\mu\nu}k_\mu k_\nu
=\frac{1}{2}g^{\mu\nu}\nabla_\mu S \nabla_\nu S. 
\end{equation}
The modification of the eikonal \eqref{eq:tildes} and 
Eq.~\eqref{eq:kdphi2} suggest the following Hamiltonian:
\begin{eqnarray}
\widetilde{\mathcal H}
=\frac{1}{2}g^{\mu\nu}(\nabla_\mu \widetilde S -\sigma\varphi_\mu) 
(\nabla_\nu \widetilde S-\sigma\varphi_\mu), 
\label{eq:Hami}
\end{eqnarray}
where we have defined 
\begin{equation}
\varphi_\mu:=\frac{1}{2}\varepsilon_{\mu\nu\rho\lambda} u^\nu  \nabla^\rho u^\lambda. 
\end{equation}
This expression was first derived by a group at Osaka City University\cite{OCUgroup} 
in a different way. 

Our aim is to modify the ordering of the field equations 
so that Eq.~\eqref{eq:Hami} is obtained. 
It will be seen in Eq.~\eqref{eq:phaseconst} that 
our procedure eliminates 
the phase shift Eq.~\eqref{eq:kdphi2}. 
We do not change the form of the eikonal ansatz~\eqref{eq:ansatz} 
but formally put the tilde `` $\widetilde ~$ " on all 
quantities, 
as follows:
\begin{equation}
A_\mu=\left(\widetilde a_\mu+\epsilon \widetilde b_\mu+\mathcal O(\epsilon^2)\right)
e^{i\widetilde S/\epsilon}. 
\label{eq:ansatz2}
\end{equation}
To obtain 
the Hamiltonian~\eqref{eq:Hami} 
we change the orders of significance 
in the geometrical optics approximation by 
rewriting the gradient operator as follows: 
\begin{equation}
\nabla_\mu\rightarrow
\nabla_\mu  - i\epsilon^{-1}\sigma\varphi_\mu
+ i\sigma\varphi_\mu. 
\label{eq:replace}
\end{equation} 
Then, the Lorenz gauge equation and wave equation become
\begin{eqnarray}
\nabla_\mu A^\mu=0&\rightarrow& (\nabla_\mu  - i\epsilon^{-1}\sigma\varphi_\mu
+ i\sigma\varphi_\mu)A^\mu=0, 
\label{eq:lorenz}
\\
\nabla_\nu\nabla^\nu
A_\mu=\mathcal O(\epsilon^0)&\rightarrow&(\nabla_\nu-i\epsilon^{-1}\sigma\varphi_\nu+i\sigma\varphi_\nu)
(\nabla^\nu-i\epsilon^{-1}\sigma\varphi^\nu+i\sigma\varphi^\nu)
A_\mu=\mathcal O(\epsilon^0). 
\label{eq:waveeq}
\end{eqnarray}
This modification is trivial if we take $\epsilon\rightarrow 1$, 
but this enhances the effect of the circular polarization to 
the leading order. 

From the order of $\epsilon^{-1}$ in Eq.~\eqref{eq:lorenz}, 
we obtain 
\begin{equation}
\widetilde a^\mu q_\mu=0,
\end{equation}
where 
\begin{equation}
q_\mu=\nabla_\mu \widetilde S-\sigma\varphi_\mu. 
\end{equation}
This means that the polarization 
vector $\widetilde a^\mu$ is perpendicular to the ray direction given by 
$q^\mu$. 

From the order of $\epsilon^{-2}$ in Eq.~\eqref{eq:waveeq}, 
we have 
\begin{equation}
q^\mu q_\mu=0. 
\label{eq:null2}
\end{equation}
This equation is identical to 
$\widetilde{\mathcal H}=0$
and 
$\widetilde{\mathcal H}$ is simply the Hamiltonian for the ray trajectory. 

In the same way as with Eq.~\eqref{eq:liek}, 
we extend the vector $q^\mu$ with the Lie transport 
along the integral curves of $\xi^\mu$, that is, 
\begin{equation}
\mathcal L_\xi q^\mu
%=\xi^\nu\nabla_\nu k^\mu-k^\nu\nabla_\nu\xi^\mu
=0. 
\end{equation}
Then, we define the frequency $\widetilde \omega$ as follows: 
\begin{equation}
\widetilde \omega:=-\xi^\mu q_\mu. 
\end{equation}
It should be noted that this frequency is not constant in general, 
in contrast to $\omega=-\xi^\mu k_\mu$. 
Since $q^\mu$ depends on the helicity $\sigma$, 
$\widetilde \omega$ also depends on $\sigma$. 
We define 
the spacelike unit vector 
along the modified ray direction 
$\widetilde n^\mu$ as follows:
\begin{equation}
\widetilde n^\mu:=\frac{\sqrt{h}}{\widetilde \omega}q^\mu-u^\mu. 
\end{equation}
Following the same procedure as in Sec.~\ref{sec:basevec}, 
we can define the modified circular polarization base vector $\widetilde m^\mu$ 
associated with $\widetilde n^\mu$.

From the order of $\epsilon^{-1}$ in Eq.~\eqref{eq:waveeq}, 
we obtain
\begin{equation}
q^\nu\nabla_\nu \widetilde a^\mu+\frac{1}{2}\widetilde a^\mu\nabla_\nu q^\nu
+i\sigma q^\nu \varphi_\nu \widetilde a^\mu=0. 
\end{equation}
We divide $\widetilde a^\mu$ into the real scalar 
amplitude $\widetilde a$ and the circular 
polarization vector 
$\widetilde \ell^\mu:=\widetilde a^\mu /a=\widetilde m^\mu\ee^{i\widetilde \varphi}$. 
Contracting with $\overline {\widetilde m}^\mu$, we obtain 
\begin{equation}
q^\nu\nabla_\nu \widetilde a+i\widetilde aq^\nu\nabla_\nu\widetilde \varphi
+\widetilde a\overline {\widetilde m}_\mu q^\nu\nabla_\nu
\widetilde m^\mu+\frac{1}{2}\widetilde a \nabla_\nu q^\nu
+i\sigma \widetilde a  q^\nu\varphi_\nu=0.
\label{eq:wave2m}
\end{equation}
Similar to Eqs.~\eqref{eq:kdphi1} and \eqref{eq:kdphi2}, 
the third term of this equation can be 
rewritten as 
\begin{eqnarray}
\overline {\widetilde m}_\mu q^\nu\nabla_\nu \widetilde m^\mu&
=&-\frac{\widetilde \omega}{\sqrt{h}}
\widetilde m_\mu(\widetilde n^\nu+u^\nu)\nabla_\nu\overline {\widetilde m}^\mu\cr
&=&-\frac{\widetilde \omega}{\sqrt{h}}
\widetilde m_\mu u^\nu\nabla_\nu\overline {\widetilde m}^\mu\cr
&=&-\frac{\widetilde \omega}{h}
\widetilde m_\mu\xi^\nu\nabla_\nu\overline {\widetilde m}^\mu\cr
&=&-\frac{\widetilde \omega}{h}
\widetilde m^\mu\overline {\widetilde m}^\nu\nabla_\nu\xi_\mu\cr
&=&-i\sigma q^\mu\varphi_\mu. 
\end{eqnarray}
Then, from the real 
and imaginary parts of Eq.~\eqref{eq:wave2m}, we obtain 
the following two equations: 
\begin{eqnarray}
&&\nabla_\mu (\widetilde a^2 q^\mu)=0, 
\label{eq:photonconv}
\\
&&q^\mu\nabla_\mu \widetilde\varphi=0. 
\label{eq:phaseconst}
\end{eqnarray}
Eq.~\eqref{eq:photonconv} describes the 
photon number conservation and 
Eq.~\eqref{eq:phaseconst} indicates that 
the phase $\widetilde\varphi$ is constant 
along the ray trajectory. 
Eq.~\eqref{eq:phaseconst} is the desired result 
for the modification. 

From the Hamiltonian \eqref{eq:Hami}, 
we obtain the following equation of motion 
for the ray trajectory:
\begin{equation}
q^\nu\nabla_\nu q^\mu=\sigma f^\mu_{~\nu} q^\nu, 
\label{eq:eom}
\end{equation}
where 
\begin{equation}
f_{\mu\nu}=\nabla_\mu \varphi_\nu-\nabla_\nu\varphi_\mu. 
\label{eq:effforce}
\end{equation}
Performing (1+3)-decomposition, we can derive 
Eq.~(112) in Ref.~\cite{Frolov:2011mh}.

We also perform the replacement \eqref{eq:replace} 
in the expression \eqref{eq:field}. 
We obtain 
\begin{equation}
F_{\mu\nu}={\rm Re}\left\{\left(\nabla_\mu-i\varepsilon^{-1}\sigma \varphi_\mu
+i\sigma\varphi_\mu\right) A_\nu-\left(\nabla_\nu-i\varepsilon^{-1}\sigma \varphi_\nu
+i\sigma\varphi_\nu\right) A_\mu\right\}
\simeq2a {\rm Re}\left\{i\ee^{i\widetilde S}q_{[\mu}\widetilde \ell_{\nu]}\right\}
\label{eq:field2}
\end{equation}
at the leading order of the modified geometrical optics approximation. %
Then, the energy momentum tensor $T^{\mu\nu}$ is given by
\begin{eqnarray}
T^{\mu\nu}
&\simeq&\frac{\widetilde a^2}{8\pi}q^\mu q^\nu
\left(1-{\rm Re}\left\{\ee^{2i\widetilde S}\widetilde \ell_\lambda\widetilde \ell^\lambda\right\}\right). 
\end{eqnarray}
Averaging over several wavelengths, we obtain
\begin{equation}
\left<T^{\mu\nu}\right>\simeq\frac{\widetilde a^2}{8\pi}q^\mu q^\nu. 
\end{equation}
This expression indicates that the 
energy flux is proportional to $q^\mu$ and null 
at the leading order of the modified geometrical optics approximation.

%%%%%%%%%%%%%%%%%%%%%%%%%%%%%%%%%%%%%%%%%%%%%%%%%%%%%%%%%%%%%%%%
\section{Gravitational Spinoptics}
\label{sec:GS}
%%%%%%%%%%%%%%%%%%%%%%%%%%%%%%%%%%%%%%%%%%%%%%%%%%%%%%%%%%%%%%%%

%%%%%%%%%%%%%%%%%%%%%%%%%%%%%%%%%%%%%%%%%%%%%%%%%%%%%%%%%%%%%%%%
\subsection{Standard Geometrical Optics}
%%%%%%%%%%%%%%%%%%%%%%%%%%%%%%%%%%%%%%%%%%%%%%%%%%%%%%%%%%%%%%%%

The eikonal ansatz for the metric perturbation is given by follows: 
\begin{equation}
h_{\mu\nu}=(a_{\mu\nu}+\epsilon b_{\mu\nu}+\mathcal O(\epsilon^2))\ee^{iS/\epsilon}.   
\end{equation}
Hereafter we work in the transverse-traceless gauge. 
The transverse gauge equation and the wave equation are given by 
\begin{eqnarray}
&&\nabla_\mu h^{\mu\nu}=0, 
\label{eq:lorenzgw1}
\\
&&\nabla_\rho\nabla^\rho
h_{\mu\nu}+2R_{\rho \mu \lambda \nu}h^{\rho \lambda}=0. 
\label{eq:waveeqgw1}
\end{eqnarray}

From the order of $\epsilon^{-1}$ in Eq.~\eqref{eq:lorenzgw1}, 
we obtain 
\begin{equation}
a_{\mu\nu} k^\nu=0,
\end{equation}
where $k_\mu$ is given by \eqref{eq:kdef}. 
From the order of $\epsilon^{-2}$ in Eq.~\eqref{eq:waveeqgw1}, 
we obtain the same equation as Eq.~\eqref{eq:null1} and 
\eqref{eq:nullgeo}. 
From the order of $\epsilon^{-1}$ in Eq.~\eqref{eq:waveeqgw1}, 
we obtain
\begin{equation}
k^\rho\nabla_\rho a_{\mu\nu}+\frac{1}{2}a_{\mu\nu}\nabla_\rho k^\rho
=0. 
\end{equation}
As in Eq.~\eqref{eq:divide1}, we divide $a_{\mu\nu}$ as follows:
\begin{equation}
a_{\mu\nu}=\alpha
\ell_{\mu\nu}~,~~\ell^{\mu\nu} \overline \ell_{\mu\nu}=1~,~~\alpha\in \mathbb R. 
\label{eq:divide2}
\end{equation}
We can then obtain the following two equations:
\begin{eqnarray}
\nabla_\rho \left(\alpha^2 k^\rho\right)&=&0, 
\label{eq:gwconv1}\\
k^\rho\nabla_\rho \ell_{\mu\nu}&=&0. 
\label{eq:paratrgw}
\end{eqnarray}
Eq.~\eqref{eq:gwconv1} describes the graviton number conservation and 
Eq.~\eqref{eq:paratrgw} indicates parallel transport of 
the polarization tensor $\ell_{\mu\nu}$ along the null geodesic 
generated by $k^\mu$. 

From Isaacson's formula\cite{Isaacson:1967zz,Isaacson:1968zza}, 
the effective energy momentum tensor for gravitational waves 
can be written as 
\begin{equation}
\left<T^{\rm (GW)}_{\mu\nu}\right>
=\frac{1}{32\pi}\left<{\rm Re}\left\{\nabla_\mu h_{\rho\lambda}\right\} 
{\rm Re}\left\{\nabla_\nu h^{\rho\lambda}\right\}\right>. 
\label{eq:tgw}
\end{equation}
At the leading order of the geometrical optics approximation, 
we obtain
\begin{equation}
\left<T^{\rm (GW)}_{\mu\nu}\right>
\simeq\frac{1}{64\pi}\alpha^2k_\mu k_\nu. 
\end{equation}
This expression indicates that the energy flux of the gravitational waves 
is proportional to 
$k^\mu$ and null 
at the leading order of the standard geometrical optics approximation.

%%%%%%%%%%%%%%%%%%%%%%%%%%%%%%%%%%%%%%%%%%%%%%%%%%%%%%%%%%%%%%%%
\subsection{Base Setting and Parallel Transport of the Polarization Tensor}
%%%%%%%%%%%%%%%%%%%%%%%%%%%%%%%%%%%%%%%%%%%%%%%%%%%%%%%%%%%%%%%%

Let us consider the base tensor fields for linear polarization tensors given by 
\begin{equation}
e_+^{\mu\nu}=\frac{1}{\sqrt{2}}\delta^{AB}e^\mu_Ae^\nu_B~,~~
e_\times^{\mu\nu}=\sqrt{2}e^{(\mu}_1e^{\nu)}_2, 
\end{equation}
where round brackets around indices denote symmetrization. 
These satisfy
\begin{equation}
g_{\mu\nu}g_{\rho\lambda}e_\times^{\mu\rho}e_\times^{\nu\lambda}=1~,~~
g_{\mu\nu}g_{\rho\lambda}e_+^{\mu\rho}e_+^{\nu\lambda}=1~,~~
g_{\mu\nu}g_{\rho\lambda}e_\times^{\mu\rho}e_+^{\nu\lambda}=0~,~~ 
%\end{equation}
%and 
%\begin{equation}
g_{\rho\lambda}e_+^{\mu\rho}e^{\lambda\nu}_\times=e^{[\mu}_1e^{\rho]}_2. 
\end{equation}
For the circular polarization specified by $\sigma$, 
we can define the polarization base tensor $m_{\mu\nu}$ by 
\begin{equation}
m^{\mu\nu}=\frac{1}{\sqrt{2}}\left(e_+^{\mu\nu}+i\sigma e_\times^{\mu\nu}\right). 
\end{equation}
This satisfies 
\begin{equation}
m_{\mu\nu}m^{\mu\nu}=1~,~~
\overline m_{\mu\nu}\overline m^{\mu\nu}=1~,~~
\overline m_{\mu\nu}m^{\mu\nu}=0~,~~ 
%\end{equation}
%and
%\begin{equation}
m^{\mu\rho}\overline m_\rho^{~\nu}=i\sigma e^{[\mu}_1e^{\nu]}_2. 
\label{eq:mmbar}
\end{equation}
Using this circular polarization base tensor, we can write 
\begin{equation}
\ell_{\mu\nu}=m_{\mu\nu} \ee^{i\psi}. 
\end{equation}

Parallel transport of the polarization vector means that
\begin{equation}
k^\rho\nabla_\rho(m_{\mu\nu} \ee^{i\psi})=0
\Leftrightarrow
m_{\mu\nu} k^\rho\nabla_\rho (\ee^{i\psi})=-\ee^{i\psi}k^\rho\nabla_\rho m_{\mu\nu}. 
\end{equation}
Contracting with $\overline m^{\mu\nu}$, 
we obtain 
\begin{eqnarray}
ik^\mu \nabla_\mu \psi &=&
m^{\mu\nu} k^\rho\nabla_\rho \overline m_{\mu\nu}. 
\end{eqnarray}
Using Eq.~\eqref{eq:n2k}, we have 
\begin{eqnarray}
ik^\mu \nabla_\mu \psi &=&
\frac{\omega}{\sqrt{h}} m^{\mu\nu} (n^\rho+u^\rho)\nabla_\rho \overline m_{\mu\nu}\cr
&=&\frac{\omega}{\sqrt{h}} m^{\mu\nu} n^\rho \nabla_\rho \overline m_{\mu\nu}
+\frac{\omega}{h}m^{\mu\nu} \xi^\rho \nabla_\rho \overline m_{\mu\nu}\cr
&=&\frac{\omega}{h}m^{\mu\nu} \xi^\rho \nabla_\rho \overline m_{\mu\nu}
\cr
&=&2\frac{\omega}{h}m^{\mu\nu} \overline m_\nu^{~\rho} \nabla_\rho \xi_\mu 
\cr
&=&2i\sigma\frac{\omega}{h} e_1^{[\mu}e^{\rho]}_2 \nabla_\rho \xi_\mu 
\cr
&=&i\sigma u_\rho k_\lambda \varepsilon^{\mu\nu\rho\lambda} \nabla_\nu u_\mu
=2ik^\mu \nabla_\mu \varphi ,
\label{eq:kdphi_gw}
\end{eqnarray}
where we have used $e^\mu_A n^\nu\nabla_\nu \overline m_{\mu\rho}=0$, 
$\mathcal L_\xi \overline m^{\mu\nu}=0$ and Eq.~\eqref{eq:mmbar}.

%%%%%%%%%%%%%%%%%%%%%%%%%%%%%%%%%%%%%%%%%%%%%%%%%%%%%%%%%%%%%%%%
\subsection{Modified Geometrical Optics}
%%%%%%%%%%%%%%%%%%%%%%%%%%%%%%%%%%%%%%%%%%%%%%%%%%%%%%%%%%%%%%%%
Eq.~\eqref{eq:kdphi_gw} suggests the Hamiltonian
\begin{eqnarray}
\widetilde{\mathcal H}_{\rm gw}
=\frac{1}{2}g^{\mu\nu}(\nabla_\mu \widetilde S_{\rm gw} -2\sigma\varphi_\mu) 
(\nabla_\nu \widetilde S_{\rm gw} -2\sigma\varphi_\mu)
\label{eq:Hamigw}
\end{eqnarray}
for the modified geometrical optics. 
The extra factor 2 compared with Eq.~\eqref{eq:Hami} 
is expected from the spin-2 nature of gravitational waves.
We do not change the eikonal ansatz for the metric perturbation:
\begin{equation}
h_{\mu\nu}=(\widetilde a_{\mu\nu}+\epsilon \widetilde b_{\mu\nu}+\mathcal O(\epsilon^2))\ee^{i\widetilde S_{\rm gw}/\epsilon}.  
\end{equation} 
As for Eqs.~\eqref{eq:lorenz} and \eqref{eq:waveeq}, 
we rewrite the transverse gauge equation and the wave equation as follows:
\begin{eqnarray}
\nabla_\mu h^{\mu\nu}=0&\rightarrow& (\nabla_\mu  - 2i\epsilon^{-1}\sigma\varphi_\mu
+ 2i\sigma\varphi_\mu)h^{\mu\nu}=0, 
\label{eq:lorenz_gw}
\\
\nabla_\rho\nabla^\rho
h_{\mu\nu}=\mathcal O(\epsilon^0)
&\rightarrow&(\nabla_\rho-2i\epsilon^{-1}\sigma\varphi_\rho
+2i\sigma\varphi_\rho)
(\nabla^\rho-2i\epsilon^{-1}\sigma\varphi^\rho+2i\sigma\varphi^\rho)
h_{\mu\nu}=\mathcal O(\epsilon^0). 
\label{eq:waveeq_gw}
\end{eqnarray}
From the order of $\epsilon^{-1}$ in Eq.~\eqref{eq:lorenz_gw}, 
we obtain 
\begin{equation}
\widetilde a_{\mu\nu} p^\nu=0,
\end{equation}
where 
\begin{equation}
p_\mu=\nabla_\mu \widetilde S_{\rm gw}-2\sigma\varphi_\mu. 
\end{equation}
From the order of $\epsilon^{-2}$ in Eq.~\eqref{eq:waveeq_gw}, 
we obtain 
\begin{equation}
p^\mu p_\mu=0. 
\label{eq:Hami2gw}
\end{equation}
This equation is identical to $\mathcal H_{\rm gw}=0$. 
In the same way as for \ref{sec:mei}, 
we impose 
\begin{equation}
\mathcal L_\xi p^\mu=0
\end{equation}
and define the frequency $\widetilde \omega_{\rm gw}$ as follows:
\begin{equation}
\widetilde \omega_{\rm gw}:=-\xi^\mu p_\mu. 
\end{equation}
Then, 
we define the spacelike unit vector 
along the modified ray direction 
$\widetilde n_{\rm gw}^\mu$ by
\begin{equation}
\widetilde n_{\rm gw}^\mu=\frac{\sqrt{h}}{\widetilde \omega_{\rm gw}}p^\mu-u^\mu.
\end{equation}
We can then obtain the modified circular polarization 
base tensor $\widetilde m_{\mu\nu}$ associated with 
$\widetilde n_{\rm gw}^\mu$. 

From the order of $\epsilon^{-1}$ in Eq.~\eqref{eq:waveeq_gw}, 
we obtain
\begin{equation}
p^\rho\nabla_\rho \widetilde a_{\mu\nu}
+\frac{1}{2}\widetilde a_{\mu\nu}\nabla_\rho p^\rho
+2i\sigma p^\rho \varphi_\rho \widetilde a_{\mu\nu}=0. 
\label{eq:transportgw}
\end{equation}
We replace $\widetilde a_{\mu\nu}$ 
by $\alpha \widetilde m_{\mu\nu} \ee^{i\widetilde \psi}$. 
Contracting Eq.~\eqref{eq:transportgw} with 
$\overline {\widetilde m}^{\mu\nu}$, we obtain 
\begin{equation}
p^\nu\nabla_\nu \alpha+i\alpha p^\nu\nabla_\nu\widetilde\psi
+\alpha \overline {\widetilde m}_{\mu\nu} 
p^\rho\nabla_\rho
\widetilde m^{\mu\nu}+\frac{1}{2}\alpha \nabla_\nu p^\nu+2i\sigma \alpha  p^\nu\varphi_\nu=0.
\label{eq:wave2m_gw}
\end{equation}
Similar to Eq.~\eqref{eq:kdphi_gw}, 
the third term of this equation can be 
rewritten as 
\begin{eqnarray}
\overline {\widetilde m}^{\mu\nu}p^\rho\nabla_\rho 
\widetilde m_{\mu\nu}&=&-\frac{\widetilde \omega_{\rm gw}}{\sqrt{h}}
\widetilde m_{\mu\nu}(\widetilde n_{\rm gw}^\rho+u^\rho)\nabla_\rho
\overline {\widetilde m}^{\mu\nu}\cr
&=&-\frac{\widetilde \omega_{\rm gw}}{\sqrt{h}}
\widetilde m_{\mu\nu} u^\rho\nabla_\rho\overline {\widetilde m}^{\mu\nu}\cr
&=&-\frac{\widetilde \omega_{\rm gw}}{h}
\widetilde m_{\mu\nu}\xi^\rho\nabla_\rho\overline {\widetilde m}^{\mu\nu}\cr
&=&-2\frac{\widetilde \omega_{\rm gw}}{h}
\widetilde m^{\mu\nu}\overline {\widetilde m}_\nu^{~\rho}\nabla_\rho\xi_\mu\cr
&=&-2i\sigma p^\mu\varphi_\mu. 
\end{eqnarray}
Then, from the real and imaginary parts of Eq.~\eqref{eq:wave2m_gw}, we obtain 
the following two equations: 
\begin{eqnarray}
&&\nabla_\mu (\alpha^2 p^\mu)=0, 
\label{eq:gravitonconv}\\
&&p^\mu\nabla_\mu \widetilde\psi=0. 
\label{eq:phaseconstgw}
\end{eqnarray}
Eq.~\eqref{eq:gravitonconv} describes  
the graviton number conservation and 
Eq.~\eqref{eq:phaseconstgw} indicates that 
the phase $\widetilde\psi$ is constant 
along the ray trajectory. 
The equation of motion 
for the ray trajectory 
can be derived from the Hamiltonian \eqref{eq:Hamigw} as follows:
\begin{equation}
p^\nu\nabla_\nu p^\mu=2\sigma f^\mu_{~\nu} p^\nu, 
\label{eq:eom_gw}
\end{equation}
where $f^\mu_{~\nu}$ is defined in Eq.~\eqref{eq:effforce}. 

We also perform the replacement \eqref{eq:replace} 
in the expression \eqref{eq:tgw},
and find that 
\begin{equation}
\left<T^{\rm (GW)}_{\mu\nu}\right>\simeq\frac{1}{64\pi}\alpha^2p^\mu p^\nu. 
\end{equation}
This expression indicates that the 
energy flux is proportional to $p^\mu$ and null 
at the leading order of the modified geometrical optics approximation. 

%%%%%%%%%%%%%%%%%%%%%%%%%%%%%%%%%%%%%%%%%%%%%%%%%%%%%%%%%%%%%%%%
\section{Summary and Discussion}
\label{sec:SD}
%%%%%%%%%%%%%%%%%%%%%%%%%%%%%%%%%%%%%%%%%%%%%%%%%%%%%%%%%%%%%%%%

Using a four-dimensional covariant description, 
we have derived the equations of the modified geometrical 
optics in a stationary spacetime, 
previously derived in Ref.~\cite{Frolov:2011mh}. 
In the modified geometrical optics, 
the three-dimensional photon trajectory 
is modified depending on the photon helicity. 
In Ref.~\cite{Frolov:2011mh}, 
the authors used a reduced form of 
Maxwell's equations on 
the three-dimensional orbit space associated with 
the stationary Killing vector field. 
In this description, 
the four-dimensional picture is not clear. 
In contrast, 
in our procedure, the null nature of the photon trajectory 
in the modified geometrical optics 
naturally appears and the energy flux is apparently null. 
We can also see that, 
in contrast to the standard geometrical optics, 
the inner product of the stationary Killing vector 
and the tangent null vector to 
the modified photon trajectory is no longer a
conserved quantity along light paths. 
This quantity is furthermore different for left and right handed photon. 
The same procedure can be easily applied 
to the case of gravitational waves and 
we found that 
an additional factor of 2 appears 
in the modification 
between the circularly polarized photon and the graviton 
because of the spin 1 and 2 nature of 
electro-magnetic waves and gravitational waves, respectively. 

It is clear that the origin of the modification is 
in the choice of the circular polarization base vector field. 
Following Ref.~\cite{Frolov:2011mh}, 
we have taken a circular polarization base vector field 
based on the Fermi transport along the 
photon trajectory projected on the three-dimensional orbit space. 
However, it is still not clear whether this choice of 
base vector field is valid for 
describing the circularly polarized photon trajectory. 
Other choices of the base vector field 
may give rise to different modifications. 
Clearly we need to justify the 
choice of base vector field based on observations. 
This could be done by comparing the 
electro-magnetic fields given by 
the modified geometrical optics with 
those given by directly solving the wave equations. 
We leave this issue to a future work.

\section*{Acknowledgements}
We thank H.~Ishihara for helpful discussions and comments. 
We also thank the anonymous referee for careful reading of the manuscript 
and fruitful comments. 
CY is supported by a Grant-in-Aid through the
Japan Society for the Promotion of Science (JSPS).

%\bibliographystyle{h-physrev5-title}
%\bibliography{../../bibfiles/spinoptics}

\begin{thebibliography}{10}

\bibitem{Frolov:2011mh}
V.~P. Frolov and A.~A. Shoom,
\newblock Phys.Rev. {\bf D84}, 044026 (2011), arXiv:1105.5629, {\em {Spinoptics
  in a stationary spacetime}}.

\bibitem{Balazs:1958zz}
N.~Balazs,
\newblock Phys.Rev. {\bf 110}, 236 (1958), {\em {Effect of a Gravitational
  Field, Due to a Rotating Body, on the Plane of Polarization of an
  Electromagnetic Wave}}.
%%CITATION = PHRVA,110,236;%%

\bibitem{Plebanski:1959ff}
J.~Plebanski,
\newblock Phys.Rev. {\bf 118}, 1396 (1959), {\em {Electromagnetic Waves in
  Gravitational Fields}}.
%%CITATION = PHRVA,118,1396;%%

\bibitem{1982GReGr..14..865F}
F.~{Fayos} and J.~{Llosa},
\newblock General Relativity and Gravitation {\bf 14}, 865 (1982), {\em
  {Gravitational effects on the polarization plane}}.

\bibitem{Ishihara:1987dv}
H.~Ishihara, M.~Takahashi, and A.~Tomimatsu,
\newblock Phys.Rev. {\bf D38}, 472 (1988), {\em {Gravitational Faraday rotation
  induced by Kerr black hole}}.
%%CITATION = PHRVA,D38,472;%%

\bibitem{1992PhRvD..46.5407C}
P.~{Carini}, L.~L. {Feng}, M.~{Li}, and R.~{Ruffini},
\newblock \prd {\bf 46}, 5407 (1992), {\em {Phase evolution of the photon in
  Kerr spacetime}}.

\bibitem{NouriZonoz:1999pb}
M.~Nouri-Zonoz,
\newblock Phys.Rev. {\bf D60}, 024013 (1999), arXiv:gr-qc/9901011, {\em
  {Gravoelectromagnetic approach to the gravitational Faraday rotation in
  stationary space-times}}.
%%CITATION = GR-QC/9901011;%%

\bibitem{Sereno:2004jx}
M.~Sereno,
\newblock Phys.Rev. {\bf D69}, 087501 (2004), arXiv:astro-ph/0401295, {\em
  {Gravitational Faraday rotation in a weak gravitational field}}.
%%CITATION = ASTRO-PH/0401295;%%

\bibitem{Mashhoon:1973zz}
B.~Mashhoon,
\newblock Phys.Rev. {\bf D7}, 2807 (1973), {\em {Scattering of Electromagnetic
  Radiation from a Black Hole}}.
%%CITATION = PHRVA,D7,2807;%%

\bibitem{Mashhoon:1974cq}
B.~Mashhoon,
\newblock Phys.Rev. {\bf D10}, 1059 (1974), {\em {Electromagnetic scattering
  from a black hole and the glory effect}}.
%%CITATION = PHRVA,D10,1059;%%

\bibitem{Mashhoon:1974}
B.~Mashhoon,
\newblock Nature (London) {\bf 250}, 316 (1974), {\em {Can Einstein's theory of
  gravitaion be tested beyond the geometrical optics limit?}}

\bibitem{Mashhoon:1975ki}
B.~Mashhoon,
\newblock Phys.Rev. {\bf D11}, 2679 (1975), {\em {Influence of Gravitation on
  the Propagation of Electromagnetic Radiation}}.
%%CITATION = PHRVA,D11,2679;%%

\bibitem{1993PhLA..173..347M}
B.~{Mashhoon},
\newblock Physics Letters A {\bf 173}, 347 (1993), {\em {On the gravitational
  analogue of Larmor's theorem}}.

\bibitem{Ramos:2006sb}
J.~Ramos and B.~Mashhoon,
\newblock Phys.Rev. {\bf D73}, 084003 (2006), arXiv:gr-qc/0601054, {\em
  {Helicity-rotation-gravity coupling for gravitational waves}}.
%%CITATION = GR-QC/0601054;%%

\bibitem{Frolov:2012zn}
V.~P. Frolov and A.~A. Shoom,
\newblock (2012), arXiv:1205.4479, {\em {Scattering of circularly polarized
  light by a rotating black hole}}.
%%CITATION = ARXIV:1205.4479;%%

\bibitem{1967ZNatA..22.1328E}
J.~{Ehlers},
\newblock Zeitschrift Naturforschung Teil A {\bf 22}, 1328 (1967), {\em {Zum
  {\"U}bergang von der Wellenoptikzur geometrischen Optik in der allgemeinen
  Relativit{\"a}tstheorie}}.

\bibitem{Misner:1974qy}
C.~W. Misner, K.~Thorne, and J.~Wheeler,
\newblock (1974), {\em {Gravitation}}.
%%CITATION = INSPIRE-95654;%%

\bibitem{OCUgroup}
A.~Masuda {\em et~al.},
\newblock private communication .

\bibitem{Isaacson:1967zz}
R.~A. Isaacson,
\newblock Phys. Rev. {\bf 166}, 1263 (1967), {\em {Gravitational Radiation in
  the Limit of High Frequency. I. The Linear Approximation and Geometrical
  Optics}}.
%%CITATION = PHRVA,166,1263;%%

\bibitem{Isaacson:1968zza}
R.~A. Isaacson,
\newblock Phys. Rev. {\bf 166}, 1272 (1968), {\em {Gravitational Radiation in
  the Limit of High Frequency. II. Nonlinear Terms and the Ef fective Stress
  Tensor}}.
%%CITATION = PHRVA,166,1272;%%

\end{thebibliography}

\end{document}